\documentclass[twocolumn,prd,aps,showpacs,floats,letterpaper,floatfix,groupedaddress]{revtex4}

\usepackage{amssymb,amsmath}
\usepackage[dvips]{graphicx}

\usepackage{dcolumn,epsfig}

\def\be{\begin{equation}}
\def\ee{\end{equation}}
\def\beq{\begin{eqnarray}}
\def\eeq{\end{eqnarray}}

\def\f{\frac}

\newlength{\sizeonefig}
\newlength{\sizetwofig}
\begin{document}

\title{Quasinormal modes and thermodynamic phase transitions}

\author{Emanuele Berti} \email{berti@wugrav.wustl.edu}
\affiliation{Jet Propulsion Laboratory, California Institute of
  Technology, Pasadena, CA 91109, USA}

\author{Vitor Cardoso}
\email{vcardoso@fisica.ist.utl.pt} 
\affiliation{Centro Multidisciplinar de Astrof\'{\i}sica - CENTRA, Dept. de
  F\'{\i}sica, Instituto Superior T\'ecnico, Av. Rovisco Pais 1, 1049-001
  Lisboa, Portugal and \\ Dept. of Physics and Astronomy The University of
  Mississippi, University, MS 38677-1848, USA}

\date{\today}

\begin{abstract}
  It has recently been suggested that scalar, Dirac and Rarita-Schwinger
  perturbations are related to thermodynamic phase transitions of charged
  (Reissner-Nordstr\"om) black holes. In this note we show that this result is
  probably a numerical coincidence, and that the conjectured correspondence
  does not straightforwardly generalize to other metrics, such as Kerr or
  Schwarzschild (anti-)de Sitter.  Our calculations do not rule out a relation
  between dynamical and thermodynamical properties of black holes, but they
  suggest that such a relation is non-trivial.
\end{abstract}


\maketitle

\section{Introduction}

Kerr-Newman black holes (BHs) in (anti-)de Sitter space are characterized by
their mass $M$, dimensionless rotation parameter $a/M$, dimensionless charge
$Q/M$ and possibly a non-zero cosmological constant $\Lambda$ (here and in the
following we adopt geometrical units, $G=c=1$, and we write Einstein's
equations as $G_{\mu\nu}=3\Lambda g_{\mu\nu}$.) 

Davies \cite{Davies:1978mf,Davies:1989ey} pointed out that BHs can undergo a
second-order phase transition in which their specific heat changes sign as
$a/M$ and/or $Q/M$ are increased.
For our considerations below, we only need to recall that the phase transition
occurs when
%
%
%
%
\be \label{RNAdS} 
\left(\f{Q}{M}\right)^2=\left(\f{9}{4}\right)^2\Lambda M^2+\f{3}{4}\,, 
\ee
for Reissner-Nordstr\"om (anti)-de Sitter BHs $(a=0)$ and
\be \label{Kerr} 
\left(\f{a}{M}\right)^2=\f{\sqrt{5}-1}{2}\,, 
\ee
for Kerr BHs $(\Lambda=Q=0)$.

BH perturbations due to external fields of different spin have been studied
for decades \cite{Chandrasekhar:1985kt}. After a transient phase, the decay of
these perturbations can be described as a superposition of damped exponentials
of the form $\exp\left(i\omega t\right)$. For a given spacetime (i.e. for a
given cosmological constant $\Lambda$) the quasinormal mode (QNM) frequencies
$\omega=\omega_R+i\omega_I$ depend only on the BH parameters $M$, $a$ and $Q$.
They are complex numbers because oscillations are damped away by the emission
of gravitational radiation, and they are usually labeled by three integers:
$\omega=\omega_{lmn}$ (see \cite{Kokkotas:1999bd} for a review, and
\cite{Berti:2004md} for a summary of the properties of the QNM spectrum.)  Two
integers $(l,~m)$ correspond to the ``angular quantum numbers'' of the
spin-weighted spheroidal harmonics used to separate the angular dependence of
the perturbations. For spherically symmetric backgrounds, such as the
Reissner-Nordstr\"om metric ($a=\Lambda=0$), perturbations are degenerate with
respect to the azimuthal index $m$, but this degeneracy is broken when $a\neq
0$. A third integer $n(=0,1,\dots)$, called the ``overtone number'', sorts the
frequencies by the magnitude of their imaginary parts.

For Reissner-Nordstr\"om BHs and $n<n_c$ ($n_c$ being some critical value that
depends on $l$ and on the spin $s$ of the perturbing field), $\omega_R$ and
$\omega_I$ are usually monotonic functions of the charge $Q/M$.  For $n=n_c$
the real part of the QNM frequency has an extremum as a function of the
charge. When $n>n_c$, both $\omega_R$ and $\omega_I$ usually become
oscillatory functions of $Q/M$
\cite{Andersson:1996xw,Berti:2003zu,Motl:2003cd,Neitzke:2003mz}. A similar
behavior has been observed for $m=0$ QNM frequencies of Kerr BHs as function
of $a/M$ \cite{Onozawa:1996ux,Berti:2003zu,Berti:2003jh} (see also
\cite{Berti:2005eb} for a generalization to Kerr-Newman BHs.)

Jing and Pan \cite{Jing:2008an} recently computed QNM frequencies for scalar
($s=0$), Dirac ($s=1/2$) and Rarita-Schwinger ($s=3/2$) perturbations of
Reissner-Nordstr\"om BHs. They noticed that the first maximum of
$M\omega_R(Q/M)$, as computed numerically, matches within $\sim 2.5\%$ the
value of $Q/M=\sqrt{3/4}\simeq 0.866$ predicted by Eq.~(\ref{RNAdS}) for
Davies' second-order phase transition.  They conjectured that this agreement
implies a connection between dynamical and thermodynamic properties of
Reissner-Nordstr\"om BHs.  In this note we show that the observed agreement is
probably a numerical coincidence, and that such a connection (if it exists)
should be non-trivial. 

Our results can be summarized as follows:


\noindent (1) The approximate agreement found in Ref.~\cite{Jing:2008an} is
usually observed for $n=n_c+1$, not for $n=n_c$. We see no compelling reason
why this mode should be singled out from the QNM spectrum as especially
relevant.

\noindent (2) For $n=n_c+1$, the numerical agreement with Davies'
thermodynamic phase transition point is significantly worse for
``electromagnetic-type'' or ``gravitational-type'' perturbations of
Reissner-Nordstr\"om BHs, which were not considered in
Ref.~\cite{Jing:2008an}.

\noindent (3) Finally, and perhaps more convincingly, we show that the
conjecture does not hold for (i) integer-spin, $m=0$ perturbations of Kerr BHs,
and (ii) integer-spin perturbations of Schwarzschild anti-de Sitter (SAdS)
BHs.


In Section~\ref{details} we present details of our calculations. In
Section~\ref{conclusions} we argue that, if a relation between classical BH
oscillations and their thermodynamic properties exists, it is not as simple
as proposed in Ref.~\cite{Jing:2008an}.


\section{Results}
\label{details}

\subsection{Reissner-Nordstr\"om black holes}


Ref.~\cite{Jing:2008an} considered scalar ($s=0$), Dirac ($s=1/2$) and
Rarita-Schwinger ($s=3/2$) perturbations of a Reissner-Nordstr\"om BH. The
results presented in their Figure 1 for $s=0$ agree with those previously
published in \cite{Berti:2005eb}. From the calculations shown in Figure 1 of
\cite{Jing:2008an} (or Figure 1 of \cite{Berti:2005eb}, where modes are
counted starting from $n=1$ rather than from $n=0$) one deduces that, for
scalar perturbations with $l=0$, $M\omega_R$ has the first extremum when
$n=n_c=0$ and $Q/M=0.962$.  Similarly, for scalar perturbations with $l=1$,
the first extremum occurs when $n=n_c=1$ and $Q/M=0.960$. Both values are
quite far from Davies' phase transition point $Q/M\simeq 0.866$.

\begin{table}[hbt] \centering \caption{\label{tab:RN} ``Critical'' charge
    $Q/M$ corresponding to the first extremum of $M\omega_R(Q/M)$ for
    Reissner-Nordstr\"om BHs. We consider only the dominant multipolar
    component of the radiation, i.e.  $l=|s|$. For each spin $s$, in the first
    row we list the critical charge for $n=n_c$ (the overtone number for which
    an extremum in $M\omega_R$ first appears); in the second row, we show the
    charge corresponding to the first extremum for $n=n_c+1$. 
  }
\begin{tabular}{ccccc}  \hline
\hline \multicolumn{1}{c}{} & \multicolumn{3}{c}{ $Q/M$}\\
\hline
Perturbation            &$l=0$    &$l=1$   &$l=2$    \\
\hline
Scalar                  &0.962 (0) &-       &-       \\
                        &0.879 (1) &-       &-       \\
\hline
Electromagnetic-type    &-         &0.962(2)&-       \\
                        &-         &0.940(3)&-       \\
\hline
Gravitational-type      &-         &-       &0.958(1)\\
                        &-         &-       &0.910(2)\\
\hline
\end{tabular}
\end{table}

For $s=0$, the maximum in $M\omega_R(Q/M)$ gets in better agreement with the
phase transition point if we consider the next overtone $n=n_c+1$, as we show
by an explicit calculation in Table \ref{tab:RN}.  For the $s=l=0$, $n=1$
mode, and accounting for their different choice of units,
Ref.~\cite{Jing:2008an} determines the ``critical'' charge to be $Q/M=0.876$.
This is in good (if not perfect) agreement with our value $Q/M=0.879$, which
has a $1.5\%$ disagreement with the phase transition point predicted by
Davies.



Ref.~\cite{Jing:2008an} did not consider gravitational ($s=2$) and
electromagnetic ($s=1$) perturbations of Reissner-Nordstr\"om BHs. These
perturbations are coupled to each other by the BH charge, but the scattering
problem can still be reduced to a pair of wave equations (see discussions in
\cite{Chandrasekhar:1985kt,Berti:2005eb,Andersson:2003fh}.) The two equations
are said to describe ``electromagnetic-type'' and ``gravitational-type''
perturbations, depending on whether they reduce to pure electromagnetic or
pure gravitational perturbations of a Schwarzschild BH in the limit $Q/M\to
0$.
In Table \ref{tab:RN} we list the critical charge for the dominant multipoles
of ``electromagnetic type'' and ``gravitational type'' perturbations. From the
tabulated values it is quite clear that the agreement with Davies' phase
transition point does not get better even if, following
Ref.~\cite{Jing:2008an}, we consider the QNM with $n=n_c+1$ rather than that
with $n=n_c$.

In our opinion, this is evidence that the conjectured correspondence is only a
numerical coincidence. The value of the charge for which $M\omega_R$ first has
an extremum is around $Q/M\simeq 0.96$ for all integer spins $s=0,1,2$, and
this is quite far from the thermodynamic phase transition point $Q/M=0.866$.
One could still argue that the dynamical and thermodynamic properties of BHs
are not simplest when one considers coupled electromagnetic and gravitational
perturbations: see e.g.  \cite{Andersson:2003fh} for a discussion.  To show
that this is not the only reason for the discrepancy with Davies' critical
point, in the remainder of this paper we show that the conjectured
correspondence does {\it not} hold for other metrics as well. In particular,
below we consider Kerr and SAdS BHs.

\subsection{Kerr black holes}

For Kerr BHs, according to Eq.~(\ref{Kerr}) the thermodynamic phase
transition corresponds to a rotation parameter $a/M\simeq 0.786$. For moderate
values of $n$, only Kerr perturbations with $m=0$ are oscillatory functions of
$a/M$ \cite{Berti:2004md,Berti:2003jh}. For this reason, in our discussion we
only consider QNM frequencies with $m=0$.

\begin{table}[hbt]
  \centering \caption{\label{tab:Kerr} Dimensionless angular momentum parameter
    $a/M$ corresponding to the first extremum of $M\omega_R(a/M)$ for Kerr
    perturbations with $m=0$. We only consider the two lowest allowed
    values of $l$: $l=|s|,|s|+1$. In parentheses we list $n_c$, the
    overtone number for which an extremum first appears.}
\begin{tabular}{ccccc}  \hline
\hline \multicolumn{1}{c}{} & \multicolumn{4}{c}{ $a/M$}\\
\hline
Perturbation       &$l=0$    &$l=1$   &$l=2$    &$l=3$    \\
\hline
Scalar             &0.819 (0)&0.898(1)&         &         \\
Electromagnetic    &-        &0.922(1)&0.925(2) &         \\
Gravitational      &-        &-       &0.947(2) &0.937(2) \\
\hline
\end{tabular}
\end{table}

The first occurrence of local extrema in $M\omega_R(a/M)$ for $m=0$ and
different values of $l$ and $s$ is listed in Table \ref{tab:Kerr}. For $n=n_c$
there is no agreement with the thermodynamic phase transition point predicted
by Davies.  As in the Reissner-Nordstr\"om case, we verified that the
agreement would not improve if we considered the next overtone $n=n_c+1$.
For example, for $s=l=m=0$ and $n=1$ the first extremum in $M\omega_R(a/M)$
occurs when $a/M=0.503$.



\subsection{Schwarzschild-(anti-)de Sitter black holes}

Setting $Q=0$ in Eq.~(\ref{RNAdS}) we get $0=(9/4)^2 \Lambda M^2+3/4$,
$r_+=3M/2$. A phase transition occurs when
$\sqrt{|\Lambda|}\,M=-2/(3\sqrt{3})$ and
%
$\sqrt{|\Lambda|}\,r_+=1/\sqrt{3}\simeq 0.577$
for negative cosmological constant (anti-de Sitter space.) For positive
$\Lambda$ (de Sitter space), there is no phase transition.
\begin{table}[htb]
  \centering \caption{\label{tab:SAdS} BH horizon radius corresponding to the
    first extremum of $\omega_R(r_+)/\sqrt{|\Lambda|}$ for SAdS BHs. The
    number in parentheses is $n_c$.}
\begin{tabular}{ccccc}  \hline
\hline \multicolumn{1}{c}{} & \multicolumn{4}{c}{ $\sqrt{|\Lambda|}\,r_+$}\\
\hline
Perturbation       &$l=0$    &$l=1$   &$l=2$    &$l=3$    \\
\hline
Scalar             &0.39 (0) &0.61(0) &0.66 (0) &0.78 (0) \\
Electromagnetic    &-        &-       &0.92 (0) &0.92 (0) \\
Odd gravitational  &-        &-       &0.75 (1) &1.26 (1) \\
Even gravitational &-        &-       &0.73 (0) &1.04 (0) \\
\hline
\end{tabular}
\end{table}

The perturbation equations for SAdS spacetimes have been worked out in the
literature, and QNM frequencies have been studied in many papers (see
e.g.~\cite{Cardoso:2001bb,Cardoso:2003cj,Berti:2003ud,Wang:2004bv}.) We used
numerical codes we developed in the past to look for extrema in the QNM
frequencies. We have two choices to make the frequency dimensionless: 

\noindent
(i) We can rescale by the BH mass and compute $M \omega_R(r_+)$. This function
has no extrema for SAdS BHs.

\noindent
(ii) We can rescale frequencies by the cosmological constant $\Lambda$. Then
$\omega_R(r_+)/\sqrt{|\Lambda|}$
does have an extremum for the values or $r_+$ listed in Table \ref{tab:SAdS}:
compare with Fig.~1 of Ref.~\cite{Berti:2003ud}. It is clear from the Table
that these extrema are strongly dependent on $l$ and $s$, and that in general
they do not agree with Davies' phase transition point. The fact that QNM
frequencies {\it do} have an extremum close to the phase transition point is
intriguing, but the physical reason for this extremum is unclear.


\section{Conclusions}
\label{conclusions}
In the last few years, remarkable relations between classical and
thermodynamic properties of black objects have been uncovered. For instance, a
correspondence between classical and thermodynamic instabilities of a large
number of black branes conjectured by Gubser and Mitra
\cite{Gubser:2000ec,Gubser:2000mm} was proved by Reall \cite{Reall:2001ag}
(see \cite{Harmark:2007md} for a review.) Manifestations of this duality are
expected to appear in the QNM spectra. Indeed, some indications that phase
transitions do show up in the QNM spectrum were provided in specific cases by
various authors \cite{Koutsoumbas:2006xj,Shen:2007xk,
Rao:2007zzb,Myung:2008ze,Koutsoumbas:2008pw}.
The main purpose of this note is to show that the correspondence is not as
simple as proposed in \cite{Jing:2008an}.

\vskip 1cm
\section*{Acknowledgements}
We thank Kostas Kokkotas, Leftheris Papantonopoulos and George Siopsis for
discussions. V.C.  acknowledges financial support from the Funda\c c\~ao para
a Ci\^encia e Tecnologia (FCT) –- Portugal through projects
PTDC/FIS/64175/2006 and POCI/FP/81915/2007. E.B.'s research was supported by
an appointment to the NASA Postdoctoral Program at the Jet Propulsion
Laboratory, California Institute of Technology, administered by Oak Ridge
Associated Universities through a contract with NASA. Copyright 2008
California Institute of Technology. Government sponsorship acknowledged.


\end{document}